\documentclass[11pt]{article}
\usepackage{geometry}                
\usepackage{graphicx}
\usepackage{amssymb}
\usepackage{amsmath}
\usepackage{amsfonts}
\usepackage{graphicx}
\usepackage{epstopdf}
\newcommand{\E}{\mathbb{E}}

\title{Eroding market stability by proliferation of financial instruments}
\author{Fabio Caccioli$^{1,2}$, Matteo Marsili$^{3} $ and Pierpaolo Vivo$^{3}$\\
{\em$^1$ SISSA, via Beirut 2-4, 34014 Trieste, Italy}\\
{\em$^2$Istituto Nazionale di Fisica Nucleare, sezione di Trieste, Via Valerio 2, 34127 Trieste, Italy}\\
{\em$^3$ The Abdus Salam International Centre for Theoretical Physics,}\\
{\em Strada Costiera 11, 34014 Trieste, Italy}
}

\begin{document}
\maketitle

\begin{abstract}
We contrast Arbitrage Pricing Theory (APT), the theoretical basis for the development of financial instruments, with a dynamical picture of an interacting market, in a simple setting. The proliferation of financial instruments apparently provides more means for risk diversification, making the market more efficient and complete. In the simple market of interacting traders discussed here, the proliferation of financial instruments erodes systemic stability and it drives the market to a critical state characterized by large susceptibility, strong fluctuations and enhanced correlations among risks.
This suggests that the hypothesis of APT may not be compatible with a stable market dynamics. In this perspective, market stability acquires the properties of a common good, which suggests that appropriate measures should be introduced in derivative markets, to preserve stability.
\end{abstract}

"In my view, derivatives are financial weapons of mass destruction, carrying dangers that, while now latent, are potentially lethal".
At the time of Warren Buffet warning \cite{Buffet} markets were remarkably stable and the creativity of financial engineers was pushing the frontiers of Arbitrage Pricing Theory to levels of unprecedented sophistication.
It took 5 years and a threefold increase in size of credit derivative markets \cite{GFSRApr}, for Warren Buffet "time bomb" to explode \cite{Buffet}.

Paradoxically, this crisis occurred in a sector which has been absorbing human capital with strong scientific and mathematical skills, as no other industry. Apparently, this potential was mostly used to increase the degree of sophistication of mathematical models, leaving our understanding lagging behind the increasing complexity of financial markets.

Recent events raise many questions:
\begin{itemize}
  \item The sub-prime mortgage market was approximately 5\% of the US real estate market, and the risk was well diversified across the whole world. Why did such a small perturbation cause so large an effect?
  \item Why are stock prices affected by a crisis in derivative markets?
  \item Why did correlations between risks grow so large during the recent crisis \cite{GFSROct}, with respect to their {\em bona-fide} estimates before the crisis?
\end{itemize}

These and other questions find no answer in the theories of mathematical economics \cite{Bouchaud}. Indeed, real markets look quite different with respect to the
picture which Arbitrage Pricing Theory (APT) \cite{pliska} gives of them. The problem is that APT is not merely a theoretical description of a phenomenon, as other theories in Natural Sciences. It is the theory on which financial engineering is based. It enters into the functioning of the system it is describing, i.e. it is part of the problem itself.

The key function of financial markets is that of allowing inter-temporal exchanges of wealth, in a state contingent manner. Taking this view, APT makes it possible to give a present monetary value to future risks, and hence to price complex derivative contracts\footnote{Part of Buffet's concerns, was related to the fact that the same mechanisms can be reversed, in a free market, to transfer and accumulate risks into the future for present return. Our focus, as we shall see, is more general and has to do with the inadequacy of the assumptions on which pricing theories are based.}. In order to do this, APT relies on concepts, such as perfect competition, market liquidity, no-arbitrage and market completeness, which allows one to neglect completely the feedback of trading on market's dynamics. These concepts are very powerful, and indeed APT has been very successful in stable market conditions. In addition, the proliferation of financial instruments provides even further instrument for hedging and pricing other derivatives. So the proliferation of financial instruments produces precisely that arbitrage-free, complete market which is hypothesized by APT\footnote{In this view, as detailed for example in Ref. \cite{MertonBodie}, market failures (e.g. crashes) are seen as arising from failures in regulatory institutions. The focus is then on the design of institutional structures which can guarantee that markets operate close to the ideal conditions. Such view has been also advocated in the context of the 2007-2008 crisis \cite{Shiller08}.}.

In theory. In practice, markets are never perfectly liquid. The very fact that information can be aggregated into prices, requires that prices respond to trading (see e.g. \cite{Osler} for evidence on FX markets or \cite{empirical2} for equity markets). In other words, it is because markets are illiquid that they can aggregate information into prices. Liquidity indeed is a matter of time scale and volume size \cite{empirical,empirical2}. This calls for a view of financial markets as interacting systems. In this view, trading strategies can affect the market in important ways\footnote{For example, Ref. \cite{pfolio} argues that even as innocent trading practices as portfolio optimization strategies, can cause dynamic instabilities if their impact on the underlying market is large enough.}. Both theoretical models and empirical research, show that trading activity implied by derivatives affects the underlying market in non-trivial ways \cite{pinning}. Furthermore, the proliferation of financial instruments (Arrow's securities), in a model with heterogeneous agents, was found to lead to market instability \cite{Hommes}.

The aim of this paper is to contrast, within a simple framework,  the picture of APT with a dynamical picture of a market as an interacting system. We show that while the introduction of derivatives makes the market more efficient,
competition between financial institutions naturally drives the market to a critical state characterized by a sharp singularity. Close to the singularity the market exhibits the three properties alluded to above: {\em 1)} a strong susceptibility to small perturbations and {\em 2)} strong fluctuations in the underlying stock market. Furthermore {\em 3)} while correlations across different derivatives is largely negligible in normal times, correlations in the derivative market are strongly enhanced in stress times, when the market is close to the critical state. In brief, this suggests that the hypothesis of APT may not be compatible with the requirement of a stable market.

Capturing the increasing complexity of financial markets in a simple mathematical framework is a daunting task. We draw inspiration from the physics of disordered systems, in physics, such as inhomogeneous alloys
which are seen as systems with random interactions.
Likewise, we characterize the typical properties of an ensemble of markets with derivatives being drawn from a given distribution. Our model is admittedly stylized and misses several important features, such as the risks associated with increased market exposure in stress conditions or the increase in demand of financial instruments in order to replicate other financial instruments. However, it provides a coherent picture of collective market phenomena.

But it is precisely because these models are simple that one is able to point out why theoretical concepts such as efficient or complete markets and competitive equilibria have non-trivial implications. The reason being that these conditions hold only in special points of the phase diagram where singularity occurs (phase transitions). It is precisely when markets approach these ideal conditions that instabilities and strong fluctuations appear \cite{JPAreview,GCMG}. Loosely speaking, this arises from the fact that the market equilibrium becomes degenerate along some directions in the phase space\footnote{The phase space is the state of the dynamical variables. The relation between fluctuations and degeneracy of equilibria has also been verified in other models, such as the Lux-Marchesi model \cite{Lux}.}. In a complete, arbitrage-free market, the introduction of a derivative contract creates a symmetry, as it introduces perfectly equivalent ways of realizing the same payoffs. Fluctuations along the direction of phase space identified by symmetries can grow unbounded. Loosely speaking, the financial industry is a factory of symmetries, which is why the proliferation of financial instruments can cause strong fluctuations and instabilities.
In this respect, the study of competitive equilibria alone can be misleading. What is mostly important is their stability with respect to adaptive behavior of agents and the dynamical fluctuations they support and generate.

The rest of the paper is organized as follows: The next section recalls the basics of APT in a simple case, setting the stage for the following sections. Then we introduce the model and discuss its behavior.
We illustrate the generic behavior of the model in a simple case where the relevant parameters are the number of different derivatives and the risk premium. In this setting we first examine the properties of competitive equilibria and then discuss the fluctuations induced within a simple adaptive process, by which banks learn to converge to these equilibria. Then we illustrate a more general model with a distribution of risk premia. This confirms the general conclusion that, as markets expand in complexity, they approach a phase transition point, as discussed above.

The final section concludes with some perspectives and suggests some measures to prevent market instability.

\section{The world of asset pricing}

A caricature of markets, from the point of view of financial engineering, is the following single period asset pricing framework \cite{pliska}:
There are only two times $t=0$ (today) and $1$ (tomorrow). The world at $t=1$ can be in any of $\Omega$ states and let $\pi^\omega$ be the probability that state $\omega=1,\ldots,\Omega$ occurs.

There are $K$ risky assets whose price is one at $t=0$ and is $1+r^\omega_k$ at $t=1$, $k=1,\ldots,K$ if state $\omega$ materializes. There is also a risk-less asset (bond) which also costs one today and pays one tomorrow, in all states\footnote{This is equivalent to considering, for the sake of simplicity, discounted prices right from the beginning.}. Prices of assets are assumed given at the outset.

Portfolios of assets can be built in order to transfer wealth from one state to the other. A portfolio $\vec{\theta}$ is a linear combination with weights $\theta_k$, $k=0,\ldots,K$ on the riskless and risky assets. The value of the portfolio at $t=0$ is
\[
V_\theta(t=0)=\sum_{k=0}^K \theta_k,
\]
which is the price the investor has to pay to buy $\vec\theta$ at $t=0$.
The return of the portfolio, i.e. the difference between its value at $t=1$ and at $t=0$, is given by
  \[
  r^\omega_\theta\equiv\sum_{k=1}^K\theta_k r_k^\omega.
  \]

The construction of Arbitrage Pricing Theory relies on the following steps

\begin{description}
  \item[No-arbitrage] It is assumed that returns $r_k^\omega$ are such that there is no portfolio $\vec{\theta}$ whose return
$r^\omega_\theta\ge 0$ is non-negative for all $\omega$ and
strictly positive on at least one state $\omega$.
  \item[Equivalent martingale measure] The absence of arbitrages implies the existence of an Equivalent Martingale Measure (EMM) $q^\omega$ which satisfies
\[
E_q[r_k]\equiv\sum_\omega q^\omega r^\omega_k=0,~~~~~\forall k=1,\ldots,K.
\]
\item[Valuation of contingent claims] For a contingent claim $f$ we mean a contract between a buyer and a seller where the seller commits to pay an amount $f^\omega$ to the buyer dependent on the state $\omega$.
If the seller can build a portfolio $\theta$ of securities such that $f^\omega=f_0+r^\omega_\theta$, i.e. which "replicates" $f$, then the seller can buy the portfolio and ensure that she can meet her commitment.
Then the value of the replicating portfolio provides a price for the contract $f$, and it is easy to see that this can be expressed in the form
\[
V_f=E_q[f]=\sum_\omega q^\omega f^\omega
\]
Claims $f$ for which this construction is possible are called marketable. If this is not possible, the parties may differ in their valuation of $f$ because of their different perception of risk. Put differently, there may be many different EMM's  consistent with the absence of arbitrage, each giving a different valuation of the contract $f$.

\item[Complete markets] If there are at least $\Omega$ independent vectors among the vectors $(r^1_k,\ldots,r^\Omega_k)$, $k=1,\ldots,K$ and $(1,\ldots,1)$, then any possible claim is marketable, which means that it can be {\em priced}. In such a case the market is called {\em complete} and the risk neutral measure is unique.
\end{description}

Summarizing, the logic of financial engineering is: assuming that markets are arbitrage free, the price of any contingent claim, no matter how exotic, can be computed. This involves some consideration of risk, as long as markets are incomplete. But if one can assumes that markets are complete, then prices can be computed in a manner which is completely independent of risk. Note indeed that the probability distribution over $\omega$ plays no role at all in the above construction. It is also worth to remark that asset return $r_k^\omega$ do not depend, by construction, on the type of portfolios which are traded in the market.

A complete, arbitrage free market is the best of all possible worlds. When markets expand in both complexity and volumes, one is tempted to argue that this is indeed a good approximation of real financial markets.

\section{A picture of the market as an interacting system}

Why are markets arbitrage free? Because, according to the standard folklore, otherwise "everybody would 'jump in' [...] affecting the prices of the security" \cite{pliska}.

Now let us assume that prices are affected not only when "everybody jumps in" but anytime someone trades, even though the effect is very small, for individual trades. In the simplified picture of the market we shall discuss below, prices depend on the balance between demand and supply; If demand is higher than supply return is positive, otherwise it is negative. Demand comes from either an exogenous state contingent process or is generated from the derivative market in order to match contracts.

Here derivatives are simply contracts which deliver a particular amount of asset in a given state (e.g. an umbrella if it rains, nothing otherwise).
Even if markets are severely incomplete, financial institutions -- which we shall call banks for short in what follows -- will match the demand for a particular derivative contract if that turns out to be profitable, i.e. if the revenue they extract from it exceeds a risk premium.

Competition in the financial industry brings on one side to smaller risk premia and on the other to a wider diversity of financial instruments being marketed in the stock market. These two effects are clearly related because the increase in financial complexity -- measured by the number of different financial instruments -- makes the market less incomplete and hence it reduces the spread in the prices of derivatives. When financial complexity is large enough the market becomes complete because each claim can be replicated by a portfolio of bond and derivatives. Beyond this point, there is an unique value for the price of each derivative, which is the one computed from APT.

Though the story is different, the conclusion seems to be the same: Efficient, arbitrage-free, complete markets. The holy grail of financial engineering.

Actually,  the road to efficient, arbitrage-free, complete markets can be plagued by singularities which arise upon increasing financial complexity. We shall illustrate this point within the stylized one period framework discussed above.

\subsection{A stylized model}

Take the one period framework as above, and let $\pi^\omega$ be the probability that state $\omega=1,\ldots,\Omega$ materializes.

Imagine there is a single risky asset\footnote{Generalization to more assets is straightforward.} and a risk-less asset. Again we shall implicitly take discounted processes, so we set the return of the risk-less security to zero.
The price of the risky asset is $1$ at $t=0$ and $1+r^\omega$, in state $\omega$. However, rather than defining at the outset the return of the asset in each state, we assume it is fixed by the law of demand and supply.

Banks develop and issue financial instruments.
In this simplified world, a financial instrument is a pair $(c,a^\omega)$ where $c$ is the cost payed to the bank at $t=0$ by the investor and $a^\omega$ is the amount of risky asset it delivers in state $\omega$ at $t=1$. We imagine there is a demand $s_0$ \footnote{More precisely, we assume investors submit a demand function which is of $s_0$ units of derivative $i$ if the price is less than $c_i$ and zero otherwise.} for each of $N$ possible financial instruments $(c_i,a_i^\omega)$, $i=1,\ldots,N$. The return of the asset at $t=1$ in each state $\omega$ is given by
\begin{equation}
\label{return}
r^\omega=d_0^\omega+\sum_{i=1}^N s_i a_i^\omega
\end{equation}
where $d_0^\omega$ is assumed to arise from investors' excess demand whereas the second term is generated by banks hedging derivative contracts\footnote{There are several ways to motivate Eq. (\ref{return}). The most transparent one, for our purposes, is that of assuming a finite liquidity of the underlying asset in a market maker setting such as that of Ref. \cite{FarmerJoshi}. Between the time ($t=0$) when derivatives are signed and the maturity ($t=1$), banks have to accumulate a position $a_i^\omega$ in the underlying for each contract $i$ they have sold ($s_i=1$). At the same time, an exogenous excess demand $d_0^\omega$ is generated by e.g. fundamental trading or other unspecified investment. The coefficient between returns and excess demand -- e.g. the market depth -- is assumed to be one here, for simplicity of notation. Market clearing between demand and supply can also be invoked to motivate Eq. (\ref{return}). Though that conforms more closely with the paradigms of mathematical economics, it introduces unnecessary conceptual complication within the stylized picture of the market given here. Likewise, we impose absence of arbitrages in the market by assuming that there exist an equivalent martingale measure $q^\omega>0$ such that $E_q[r]=0$. For the results presented below, this is an irrelevant assumption, so we shall not insist more on this.}.

The supply $s_i$ of instrument $i$ is fixed by banks in order to maximize expected profit. The expected profit for a unit supply ($s_i=1$) is given by:
\begin{equation}
\label{utility} u_i = c_i -\sum_{\omega=1}^\Omega \pi^\omega a_i^\omega(1+r^\omega).
\end{equation}
Notice that, even though $r^\omega$ depends on the supply $s_i$ of derivative $i$, by Eq. (\ref{return}), banks in a competitive market neglect this dependence and take prices (i.e. returns) as given. Rather, banks will sell derivatives if the expected profit is large enough, and will not sell it otherwise. Specifically, we assume that banks match the demand of investors (i.e. $s_i=s_0$) if $u_i>\bar u_i$, whereas $s_i=0$ if $u_i<\bar u_i$. When $u_i=\bar u_i$ then the supply can take any value $s_i\in [0,s_0]$.
Here $\bar u_i$ can be interpreted as a risk premium related to the risk aversion of banks. Indeed, as long as the market is incomplete, there is not an unique price for derivatives. Even if the prices $c_i$ were fixed on the basis of an equivalent martingale measure $q^\omega$, as in Section \ref{dynpricing}, it is not possible, for banks, to eliminate risk altogether\footnote{We do not consider here the possibility for banks to buy a portfolio of derivatives from other banks in order to hedge the derivative contracts they sell. This aspect has been addressed in a simpler setting in Ref. \cite{spiral}.}.

\subsection{A typical large complex market}

Financial markets are quite complex, with all sorts of complicated financial instruments. This situation is reproduced in our framework by assuming that demand $d_0^\omega$ and derivatives $a_i^\omega$ are drawn independently at random. Furthermore we take the limit $\Omega\to\infty$. This situation clearly defies analytical approaches for a specific realization. Remarkably, however, it is possible to characterize the statistical behavior of typical realizations of such large complex markets.

In order to do that, some comments on the scaling  of different quantities, with the number of states $\Omega$, is in order. The interesting region is the one where the number of derivatives $N$ is of the same order of the number of states $\Omega$ ($N\sim\Omega$). In this regime we expect the market to approach completeness. For this reason we introduce the variable $n=N/\Omega$. Assuming that $r^\omega$ be a finite random variable in the limit $\Omega,N\to\infty$, requires $d_0^\omega$ to be a finite random variable -- e.g. normal with mean $\bar d$ and variance $\Delta$. Likewise, we shall take $a_i^\omega$ as random variables with zero average and variance $1/\Omega$. Indeed, the second term of Eq. (\ref{return}) with
$a_i^\omega\sim 1/\sqrt{\Omega}$ is of the order of $\sqrt{N/\Omega}$, which is finite in the limit we are considering\footnote{Equivalently, one could assume $a_i^\omega$ to be of order one, but introduce a coefficient $\lambda=1/\sqrt{\Omega}$ in Eq. (\ref{return}) as a finite market depth.}.

It is convenient, in the following discussion, to introduce the parameter
\begin{equation}
\label{epsilon}
\frac{\epsilon_i}{\Omega}\equiv \bar u_i-\left[c_i-c_i^{(0)}\right].
\end{equation}
where $c_i^{(0)}= \sum_\omega \pi^\omega a_i^\omega$ is the expected price of instrument $i$, at $t=0$. The dependence on $\Omega$ in the equation above is motivated by the fact that the variance of $c_i^{(0)}$ and of the second term in Eq. (\ref{utility}), is of order ${\pi^\omega}^2\sim 1/\Omega^2$. This implies that the relevant scale for the r.h.s. of Eq. (\ref{epsilon}) is of order $1/\Omega$. The parameter $\epsilon_i$ encodes the risk premium which banks for selling derivative $i$.

In the next section, we take the simplifying assumption that $\epsilon_i=\epsilon$ does not depend on $i$, in order to illustrate the generic behavior of the model. In the next section, we shall discuss the case where $\epsilon_i$ depends on $i$.
Actually, $\epsilon_i$ encodes the risk premium that banks require for trading derivatives and hence should depend on the volatility of the realized returns $r^\omega$ and on the degree of market completeness. While keeping this in mind, we will consider $\epsilon_i$ as an independent parameter with the idea that it decreases from positive values when the volatility of the market decreases or on the route to market completeness. We shall return to this point below.

\subsection{Competitive market equilibrium}

We consider, in this section, the situation where banks chose the supply $s_i$ so as to maximize their profit, considering returns $r^\omega$ as given. The statistical properties corresponding to this state -- the so-called competitive equilibrium -- can be related to those of the minima of a global function $H$, by the following statement:

{\em The competitive market equilibrium is given by the solution of the minimization of the function}
\begin{equation}
\label{H}
H=\frac{1}{2}\sum_{\omega=1}^\Omega\pi^\omega\left(r^\omega\right)^2+\frac{\epsilon}{\Omega}\sum_{i=1}^N{s_i}
\end{equation}
{\em over the variables $0\le s_i\le s_0$, where $r^\omega$ is given in terms of $s_i$ by Eq. (\ref{return}). }


Notice that, if $\epsilon$ is negligible, this results states that a consequence of banks maximizing their utility, is that return's volatility -- the first term of Eq.(\ref{H}) -- is reduced. Actually, with $\epsilon>0$, only those derivatives which decrease volatility are traded ($s_i>0$).  This is an "unintended consequence" of banks' profit seeking behavior, i.e. a feature which emerges without agents aiming at it.

A proof of the statement above follows by direct inspection of the first order conditions and the observation that $\frac{\partial H}{\partial s_i}=-(u_i-\bar u_i)$. We first observe that $H$ as a function of any of the $s_i$ is a convex function,
\[
\frac{\partial^2 H}{\partial s_i^2}=\sum_{\omega=1}^\Omega \pi^\omega \left(a_i^\omega\right)^2\ge 0
\]
hence it has a single minimum, for $0\le s_i\le s_0$. Imagine $\{s_i^*\}$ to be the minimum of $H$. If $s_i^*=0$ it must be that
\[
\left.\frac{\partial H}{\partial s_i}\right|_{s_i=0}=\bar u_i-u_i>0.
\]
If $u_i<\bar u_i$, it is not convenient for banks to sell derivative $i$, which is consistent with zero supply ($s_i^*=0$). Likewise, if $s_i^*=s_0$ the first order condition yields $u_i>\bar u_i$, which is the condition under which banks should sell as many derivatives as possible. If instead  $0<s_i^*<s_0$, then $u_i=\bar u_i$ which is consistent with perfect competition among banks\footnote{This result can be generalized  in a straightforward way to any monotonous investors' demand function for derivatives. This goes beyond the scope of the present paper.}.

From a purely mathematical point of view, we notice that $H$ is a quadratic form which depends on $N$ variables $\{s_i\}$, through the $\Omega$ linear combinations given by the returns $r^\omega$. It is intuitively clear that, as $N$ increases, the minimum of $H$ becomes more and more shallow and, for large $N$, it is likely that there will be directions in the space of $\{s_i\}$ (i.e. linear combinations of the $s_i$) along which $H$ is almost flat. Such quasi-degeneracy of the equilibria corresponds to the emergence of symmetries discussed in the introduction. The location of the equilibrium is likely to be very sensitive to perturbation along these directions. These qualitative conclusions can be put on firmer grounds by the theoretical approach which we discuss next.

\begin{figure}[htbp]
   \centering
   \includegraphics[width=9cm]{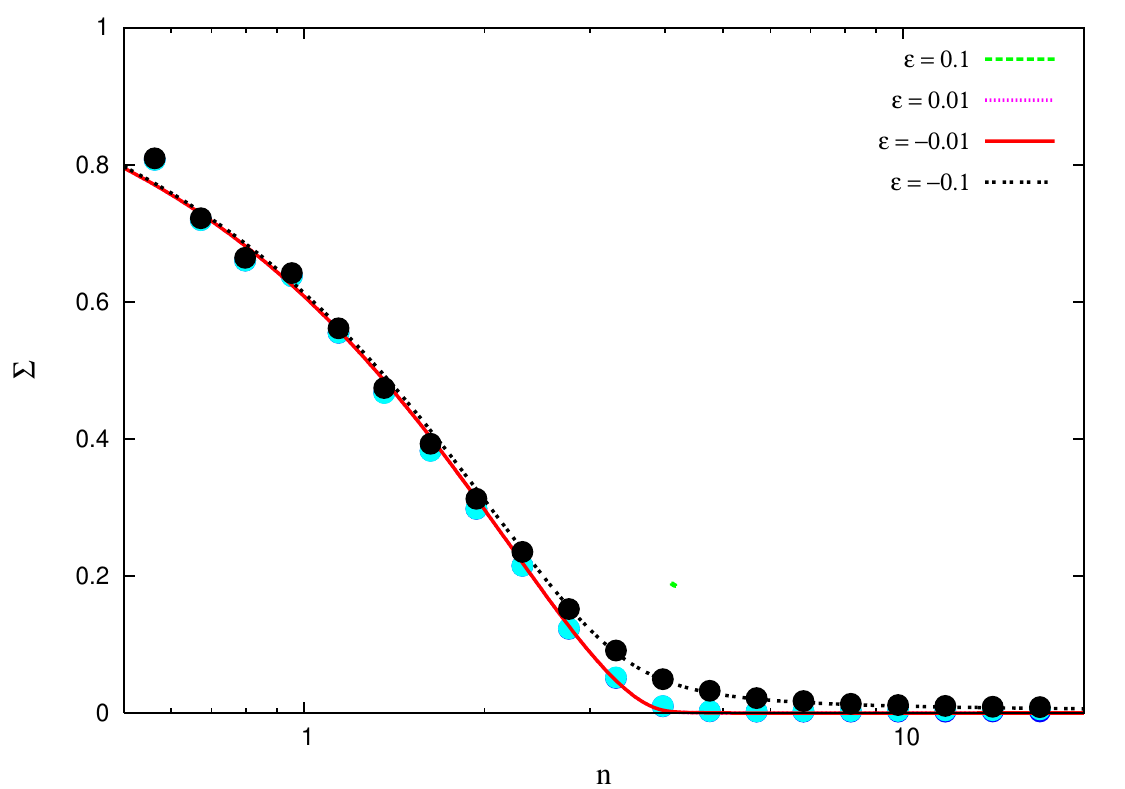} 
   \caption{Volatility $\Sigma=E_\pi\left[(r-E_\pi[r])^2\right]$ in competitive equilibria (full lines), as a function of $n=N/\Omega$, for different values of $\epsilon$. Points refer to the variance of $\bar r^\omega$ computed in numerical simulations of a system with $\Omega=64$ ($s_0=1$).}
   \label{figh}
\end{figure}

A full characterization of the solution of the minimization of $H$, in the limit $\Omega,N\to\infty$ with $N/\Omega=n$ fixed, can be derived with tools of statistical mechanics of disordered systems (see appendix), following similar lines of those in Ref. \cite{JPAreview}. Indeed, mathematically the model is quite similar to the Grand Canonical Minority Game studied in Ref. \cite{GCMG}. The solution can be summarized in the following "representative" derivative problem: Given a normal random variable $z$ with mean zero and unit variance, the supply of the "representative" derivative is given by
\begin{equation}
\label{sp0}
s_z=\max\left\{s_0,\min\left\{0,\sqrt{g+\Delta}z-\epsilon(1+\chi)\right\}\right\}
\end{equation}
where
\begin{eqnarray}
g & = & nE_z[s^2_z] \label{sp1}\\
\chi & = & \frac{nE_z[s_zz]}{ \sqrt{g+\Delta}-nE_z[s_zz]}\label{sp2}
\end{eqnarray}
are determined self-consistently in terms of expected values $E_z[\ldots]$ over functions of the variable $z$. Loosely speaking, $z$ embodies the interaction through the market of the "representative" derivative with all other derivatives. Any quantity, such as the average supply of derivatives
\begin{equation}
\bar s\equiv\lim_{\Omega\to\infty}\frac{1}{N}\sum_{i=1}^Ns_i^*=E_z[s_z]
\end{equation}
can be computed from the solution.

Figure \ref{figh} plots the volatility $\Sigma=E_\pi\left[(r-E_\pi[r])^2\right]$ as a function of $n$ for small values of $\epsilon$. As it can be seen, as the market grows in financial complexity, fluctuations in returns of the risky asset decrease and, beyond a value $n^*\simeq 4.14542\ldots$, they become very small (of order $\epsilon^2$, for small $\epsilon$). The expected return $E_\pi[r]=\bar d/(1+\chi)$ also decreases, keeping a bounded Sharpe ratio $E_\pi[r]/\sqrt{\Sigma}=\bar d/\sqrt{g+\Delta}$.

\begin{figure}[htbp]
   \centering
   \includegraphics[width=9cm]{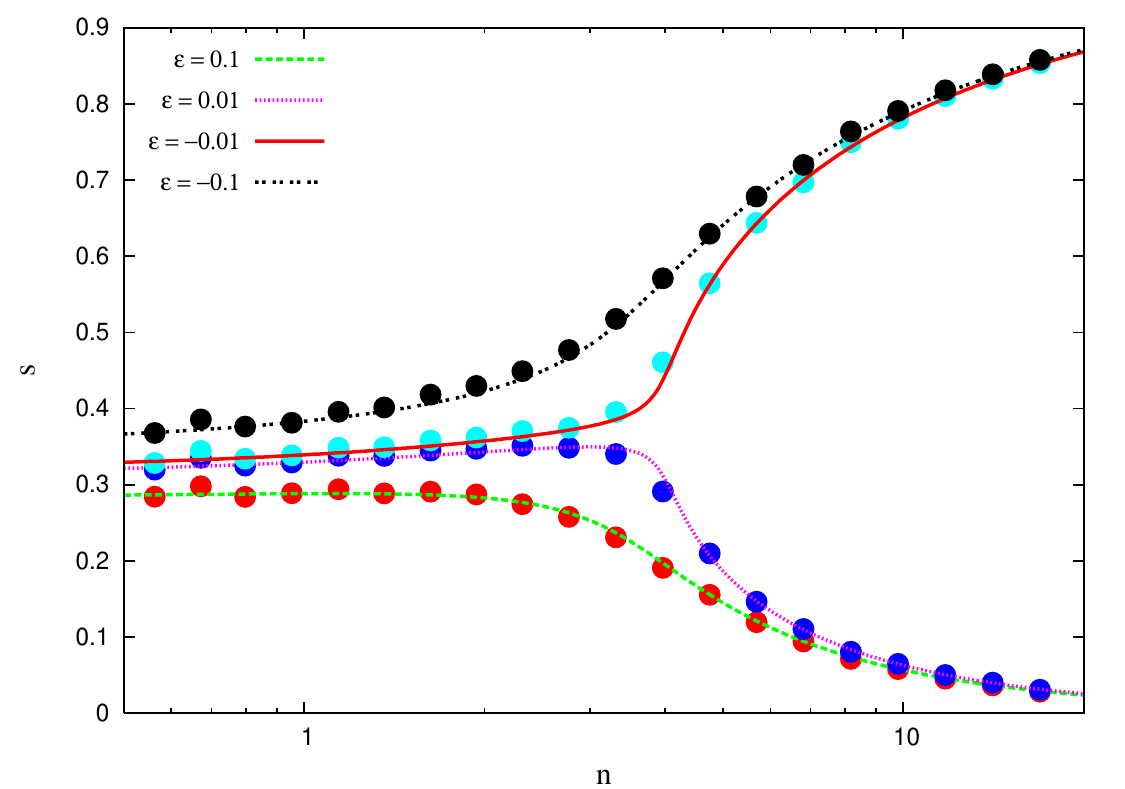} 
   \caption{Average supply $\bar s$ in competitive equilibria (full lines) for different values of $\epsilon$. Points refer to the variance of $\bar r^\omega$ computed in numerical simulations of a system with $\Omega=64$ ($s_0=1$).}
   \label{figs}
\end{figure}

This situation in the region $n>n^*$ and $\epsilon\ll 1$ resembles closely the picture of an efficient, arbitrage free complete market. Unfortunately this is also the locus of a sharp discontinuity, as shown in Fig. \ref{figs}. This plots $\bar s$ as a function of $n$ for different values of $\epsilon$. As it can be seen, as the market grows in financial complexity, it passes from a regime ($n<n^*$) where its behavior is continuous and smooth with $\epsilon$ to one ($n>n^*$) which features a sharp discontinuity at $\epsilon=0$\footnote{It should be noted that $N$ is not the actual value of derivatives traded, but the number of derivatives for which there is a demand. The number of derivatives with $s_i>0$, is well approximated by $N {\rm Prob}\{s_z>0\}$, which is less than $\Omega$, for $\epsilon>0$.}.

In particular, the discontinuity manifests clearly in the limit $s_0\to\infty$ when the demand of investors is unbounded. Then while for $\epsilon>0$ the average supply $\bar s$ is finite, for $\epsilon<0$ and $n>2$ the supply $\bar s\to\infty$ diverges as $s_0\to\infty$. In other words, while in one region ($\epsilon>0$ and $\epsilon<0$ for $n<2$) the equilibrium is controlled by the supply side, in the other region  ($\epsilon<0$, $n>2$) the equilibrium is dominated by the demand\begin{flushleft}

\end{flushleft} side. This distinction, as we shall see in the next section, applies to a more general model where $\epsilon_i$ depends on $i$.

The consequences of the singularity at $n>n^*$ and $\epsilon=0$ are not entirely evident in the competitive equilibrium setting. Indeed they manifest strongly in response functions: The market behavior close to the singularity is quite sensitive to small perturbations. For example a small change in the risk perception of banks (i.e. in $\epsilon$) can provoke a dramatic change in the volume of trading in the derivative market.

The effects of this increased susceptibility appears dramatically in the case where the interaction is repeated over time and banks learn and adapt to investors' demand. This setting not only allows us to understand under what conditions will banks learn to converge to the competitive equilibrium, but it also sheds light on the emergent fluctuation phenomena.

\subsection{Adaptive behavior of banks}

Let us assume that the context outlined above is repeated for many periods (e.g. days), indexed by $t=1,\ldots$. Let $\omega(t)$ be the state of the market at time $t$ and assume this is drawn independently from the distribution $\pi^\omega$ in each period. Accordingly, the returns $r^{\omega(t)}$ are still determined by Eq. (\ref{return}), with $\omega=\omega(t)$ and $s_i=s_i(t)$, the supply of instruments of type $i$ in period $t$.

In order to determine the latter, banks estimate the profitability of instrument $i$ on historical data. They assign a score (or attraction) $U_i(t)$ to each instrument $i$, which they update in the following manner:
\begin{equation}
\label{dynU}
U_i(t+1)=U_i(t)+u_i(t)-\bar u_i=U_i(t)-a_i^{\omega(t)}r^{\omega(t)}-\frac{\epsilon}{\Omega}.
\end{equation}
Notice that, by Eq. (\ref{utility}) and (\ref{epsilon}), $U_i(t)$ increases (decreases) if $u_i>\bar u_i$ ($u_i<\bar u_i$).
Banks supply instrument $i$ according to the simple rule
\begin{equation}
\label{sit}
s_i(t)=\left\{\begin{array}{cc}0 & \hbox{if}~~U_i(t)\le 0 \\s_0 & \hbox{if}~~U_i(t)> 0\end{array}\right.
\end{equation}
Therefore, if instrument $i$ provides an expected utility larger than the margin $\bar u_i$, its score will increase and the bank will sell it more likely. Conversely, an instrument with $u_i(t)<\bar u_i$, on average, has a decreasing score and it will not be offered by banks.

It can be checked that, in the stationary state, the average probability $s_i={\rm Prob}\{s_i(t)=1\}$ that banks issue derivative $i$ is again given by the configuration $\{s_i\}$ which minimizes $H$ in Eq. (\ref{H}). Indeed, Figs. (\ref{figh},\ref{figs}) show that averages in the stationary state of the process nicely follow the theoretical results derived above.

\begin{figure}[htbp]
   \centering
   \includegraphics[width=9cm]{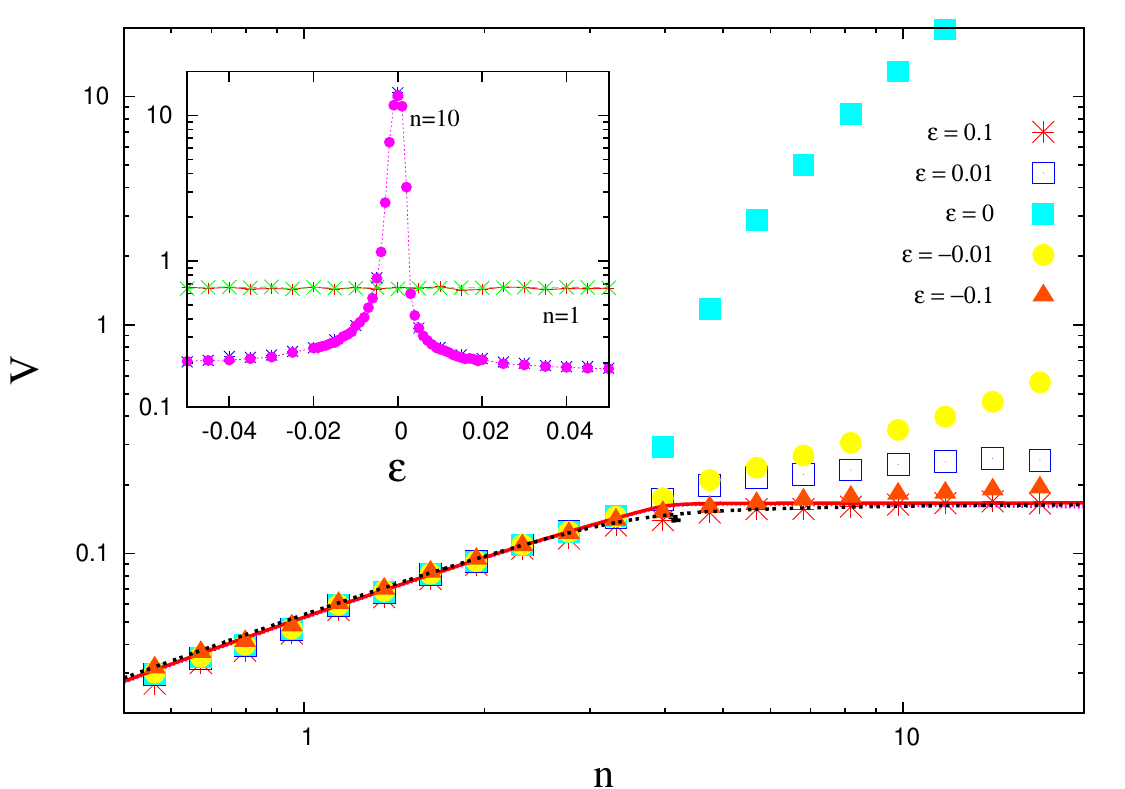} 
   \caption{Dynamical contribution to the volatility $V=E_\pi[\delta r^2]$ in numerical simulations of a system with $\Omega=64$ for different values of $\epsilon$ (points). Lines refer to the theoretical prediction in the approximation of independent variables $s_i(t)$. Inset: Total volatility $\sigma_{\bar r}^2+V$ in numerical simulations for $n=1$ ($+$ for $\Omega=128$  and $\times$ for $256$) and $n=10$ ($\ast$ for $\Omega=32$ and $\bullet$ for $\Omega=64$) as a function of $\epsilon$.}
   \label{figv}
\end{figure}

At odds with the competitive equilibrium setting, in the present case the volatility of returns also acquires a contribution from the fluctuations of the variables $s_i(t)$, which is induced by the random choice of the state $\omega(t)$. Hence, we can distinguish two sources of fluctuations in returns
\begin{equation}
r(t)=\bar r^{\omega(t)}+\delta r(t)
\end{equation}
the first depending on the stochastic realization of the state $\omega(t)$, the second from fluctuations in the learning dynamics.
Here $\bar r^\omega$ is the average return of the asset in state $\omega$, and it can be shown that it converges to its competitive equilibrium value, $\forall\omega$. Indeed the contribution of $\bar r^\omega$ to the volatility $\Sigma=E_\pi\left[(\bar r-E_\pi[\bar r])^2\right]$ closely follows the theoretical curve in Fig. (\ref{figh}).

The dynamical contribution to the volatility $V=E_\pi[\delta r^2]$ instead shows a singular behavior which reflects the discontinuity of $\bar s$ across the line $\epsilon=0$ for $n>n^*$. Our theoretical approach also allows us to estimate this contribution to the fluctuations under the assumption that the variables $s_i(t)$ are statistically independent. As Fig. \ref{figv} shows, this theory reproduces remarkably well the results of numerical simulations for $n<n^*$ but it fails dramatically for $n>n^*$ in the region close to $\epsilon=0$. The same effect arises in Minority Games \cite{JPAreview,Tobin}, where its origin has been traced back to the assumption of independence of the variables $s_i(t)$.
This suggests that in this region, the supplies of different derivatives are strongly correlated. This effect has a purely dynamical origin and it is reminiscent of the emergence of persistent correlations arising from trading in the single asset model of  Ref. \cite{WyartBouchaud}.

We have considered $\epsilon$ as a fixed parameter up to now. Actually,
in a more refined model, $\epsilon_i$ should depend on $i$ and it should be fixed endogenously in terms of the incentives of investors and banks. At the level of the discussion given thus far, it is sufficient to say that the scale of incentives of banks and investors, and hence of $\epsilon$, is fixed by the average return $\bar r=\bar d/(1+\chi)$ or by the volatility $\sqrt{\Sigma+V}$.
Both these quantities decrease as $n$ increases, so it is reasonable to conclude that $\epsilon$ should be a decreasing function of $n$, in any model where it is fixed endogenously. In other words, as the financial complexity ($n$) increases, the market should follow a trajectory in  the parameter space $(n,\epsilon)$, which approaches the critical line $n>n^*$, $\epsilon=0$.

In the next section, we address this problem in an approximate manner, extending our analysis to the case where $\epsilon_i$ depends on $i$.

\section{Asset dependent risk premia}

The model considered above can be improved in many ways.
For example, the price of derivatives can be determined according to APT, thus making the market arbitrage free both in the underlying and in derivatives. In this section we will explore this direction in the simplest possible case where the equivalent martingale measure is given. In particular we will show that  this dynamical pricing mechanism is somehow equivalent to an asset dependent risk premia framework. In order to get some insights, we will then generalize the model considered in the previous sections to the case of a quenched distribution of risk premia. This assumes that derivative prices change over much longer time-scales than those of trading.

\subsection{Dynamical pricing of derivatives}
\label{dynpricing}

In the previous sections prices of derivatives were fixed by the exogenous parameter $\epsilon$ and no role was
explicitly played by the risk neutral measure $q$. In this section we consider the case
where prices of derivatives are directly computed according to APT. The framework is the same as before, but prices  are computed through the risk neutral measure $q^{\omega}$ as
\begin{equation}
c_i=\mathrm{E}_q[a_i^\omega(1+r^\omega)]=\sum_\omega q^\omega a_i^\omega(1+r^\omega)
\end{equation}
instead of being fixed by the external parameter $\epsilon$. As before, the EMM $q$ is
fixed exogenously.  Again the
return of the risky asset is assumed to depend both on an exogenous demand and on the demand induced by the
trading of derivatives
\begin{equation}
r^{\omega}=d_0^{\omega}+\sum_{i=1}^N s_i a_i^{\omega}.
\end{equation}
The supply of derivatives $s_i\in[0,s_0]$ is then fixed by banks in order to maximize the utility function
\begin{equation}
u_i=c_i-E_\pi[a_i^\omega(1+r^\omega)]= \sum_\omega (q^\omega -\pi^\omega)
a_i^{\omega}(1+r^\omega).
\end{equation}
As in Eq. (\ref{utility}), the first term is the price of the derivative they cash from clients and the second is the expected cost of buying the stock. Again, banks will sell derivative $i$ only if $u_i>\bar u$, where $\bar u$ is a risk premium. The analog of Eq. (\ref{epsilon}) then becomes

\begin{equation}
\frac{\epsilon_i}{\Omega}=\bar u - \sum_\omega (q^{\omega} -\pi^{\omega}) a_i^{\omega}-\sum_\omega q^{\omega}
a_i^{\omega}r^{\omega}.
\end{equation}
The appearance of asset-dependent $\epsilon_i$
calls for an in-depth comparison with the simplified version studied in the previous sections.


\begin{center}
\begin{figure}
\begin{center}
\includegraphics[width=10cm]{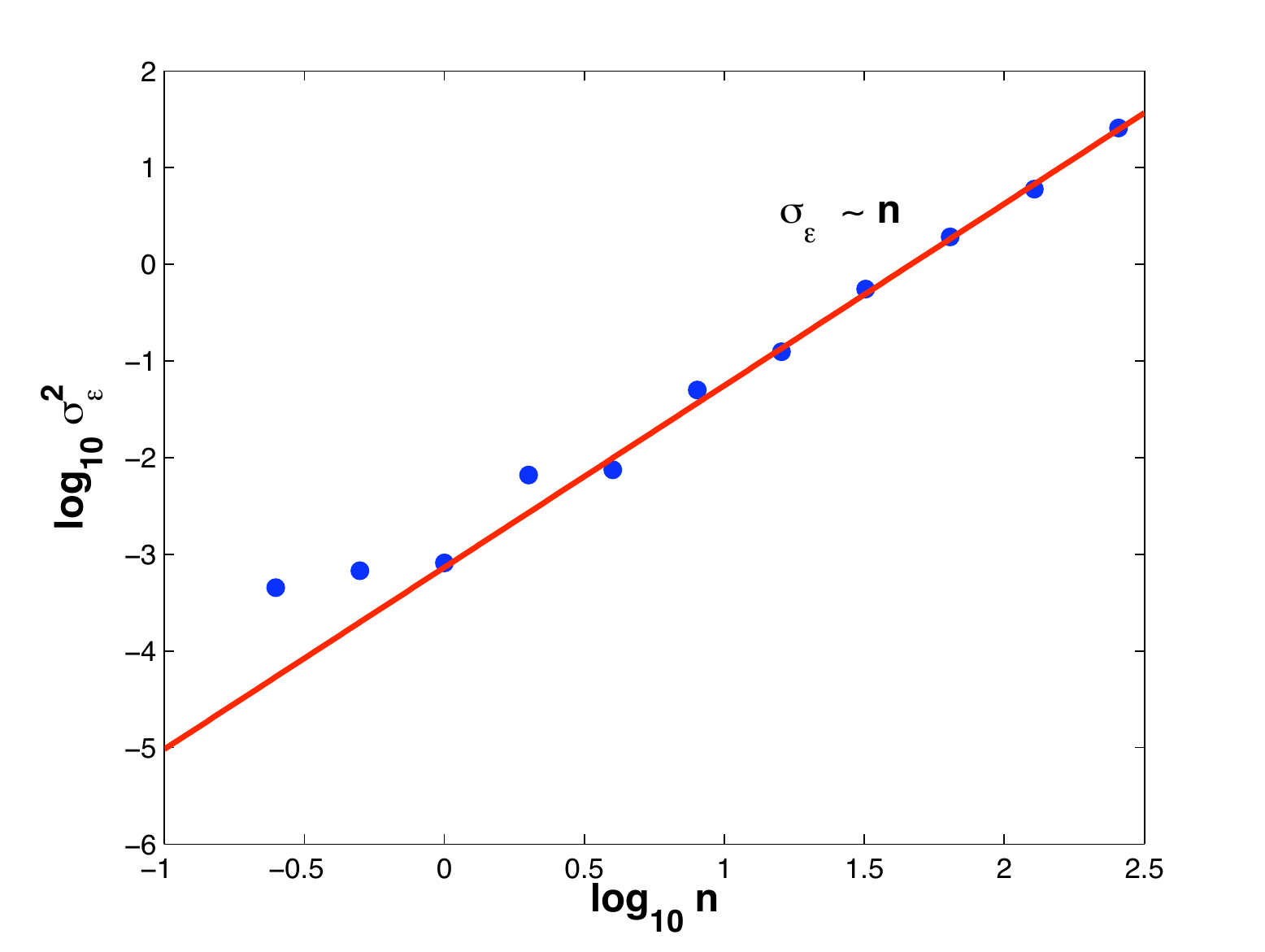}
\caption{\footnotesize {\textit{Variance of effective risk premium for a system with $\Omega=64$. } }} \label{eps_eff}
\end{center}
\end{figure}
\end{center}\begin{center}

\begin{figure}
\begin{center}
\includegraphics[width=10cm]{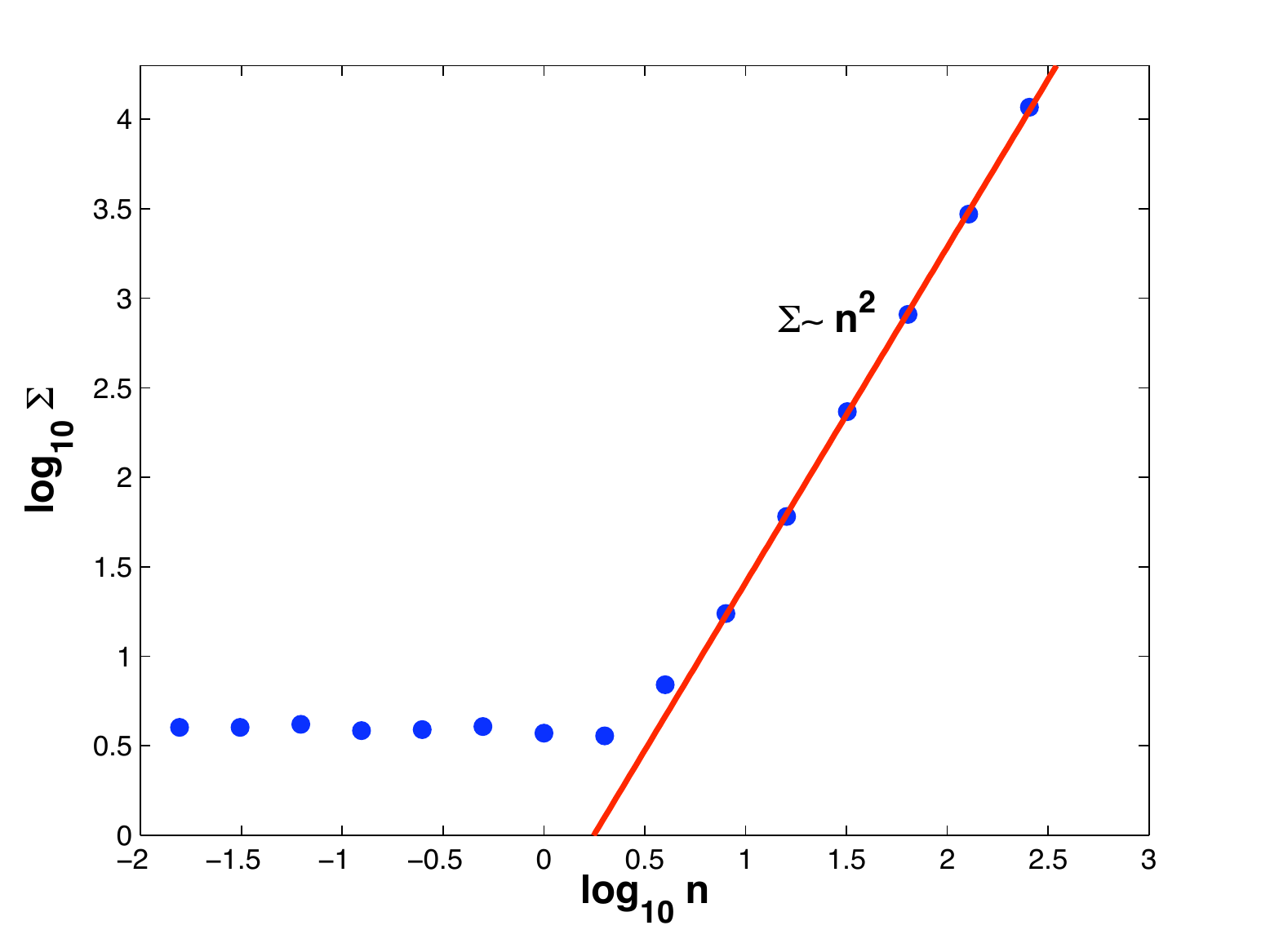}
\caption{\footnotesize {\textit{Volatility for a system with $\Omega=64$. } }} \label{volB}
\end{center}
\end{figure}
\end{center}

We performed numerical simulations using the same learning scheme
as before (Eq. \ref{dynU}), in order to detect typical features of the distribution of $\{\epsilon_i\}$. We observe that the
width of the distribution increases with $n$. A plot of the variance $\sigma_{\epsilon}^2$ of the distribution as a function of $n$ is shown in figure \ref{eps_eff}. Notice that, for $n$ large enough, $\sigma_{\epsilon}^2 \sim n^{2}$.
Figure (\ref{volB}) shows the volatility $\Sigma$ as a function of $n$. As we can see, also the volatility increases as the square of the market complexity $\Sigma\sim n^2$, so that $\Sigma\sim\sigma_{\epsilon}^2$.
As APT naturally leads to the appearence of a distribution of asset risk premia, it is instructive to generalize the above picture to the case of \emph{quenched} asset dependent risk premia, where a full analytical solution can be obtained.

\subsection{Quenched asset dependent risk premia}

In this section, we consider the case where
%

the $\{\epsilon_i\}$are taken as quenched asset-dependent random variables drawn from a gaussian
distribution with mean $\overline{\epsilon}$ and variance $\sigma_{\epsilon}^2$. In analogy with the previous
treatment, the competitive equilibrium is equivalent to finding
the ground state of the Hamiltonian
\begin{equation}\label{Ham_dis}
H=\frac{1}{2}\sum_{\omega}\pi^{\omega}(r^{\omega})^2+\sum_i\frac{\epsilon_i}{\Omega} s_i.
\end{equation}
As before, the solution can be summarized in a representative derivative problem.

Given a normal random variable $z$ with $0$ mean and unit variance, the
supply of the representative derivative is given by
\begin{equation}
s_z=    \min\left\{s_0,\max\left\{0,\frac{\sqrt{\frac{g+\Delta}{(1+\chi)^2}+\sigma_{\epsilon}^2}z-\overline{\epsilon}}{\nu} \right\}\right\},
\end{equation}
where
\begin{equation}
g=n\rm{E}_z[s_z^2],
\end{equation}
\begin{equation}
\chi=\frac{n\rm{E}_z[s_zz](1+\chi)}{\sqrt{g+\Delta+\sigma_{\epsilon}^2(1+\chi)^2}}
\end{equation}
are determined self-consistently.
\begin{center}
\begin{figure}
\begin{center}
\includegraphics[width=10cm]{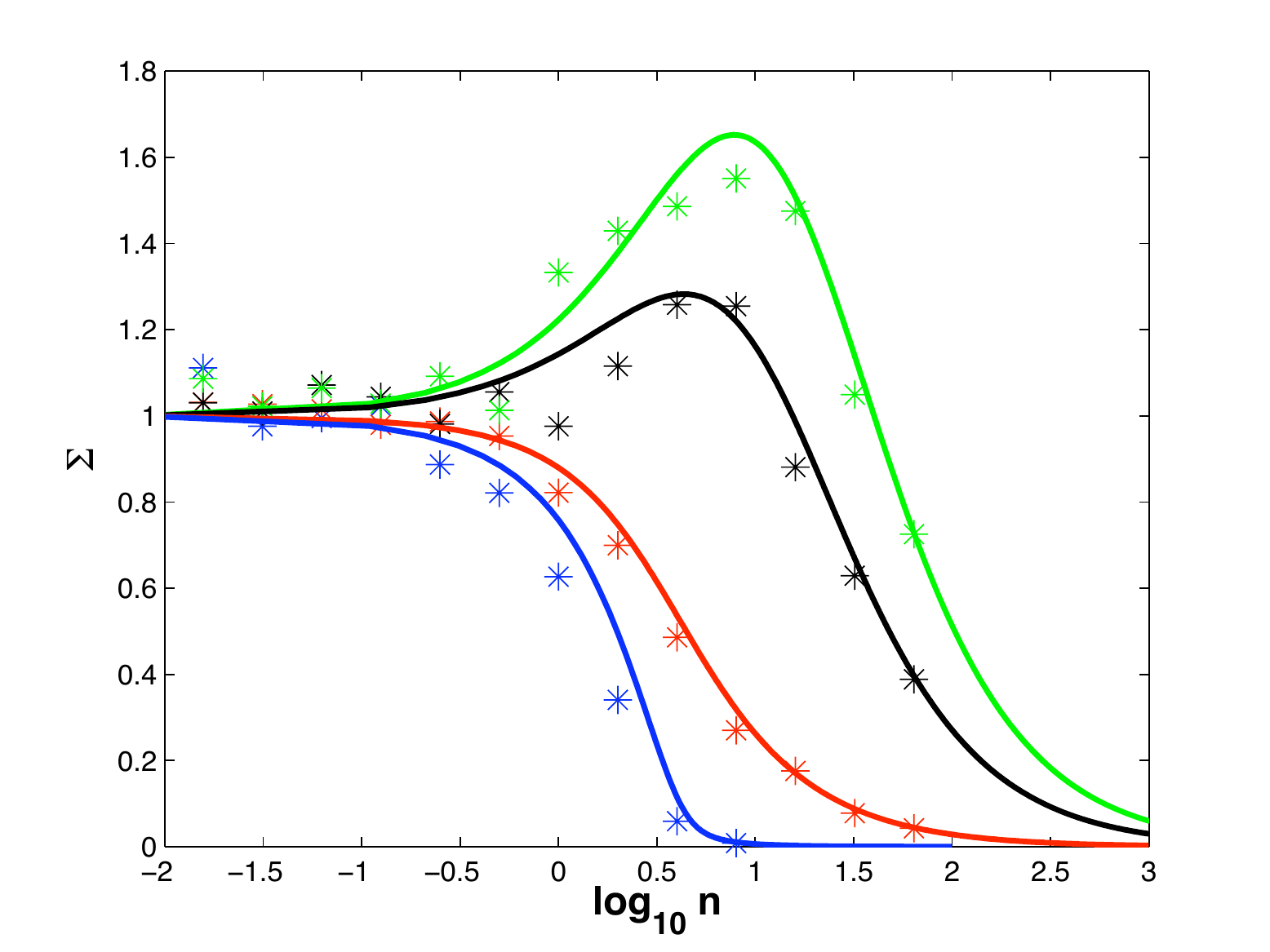}
\caption{\footnotesize {\textit{Volatility as a function of $n$ for different values of $\sigma_{\epsilon}^2$ and $\bar\epsilon=0.1$ ($s_0=1$). From top to bottom $\sigma_{\epsilon}^2=20,10,1$ and $0.01$.} }}
\label{fig:vol}
\end{center}
\end{figure}
\end{center}

\begin{figure}
\begin{center}
\includegraphics[width=10cm]{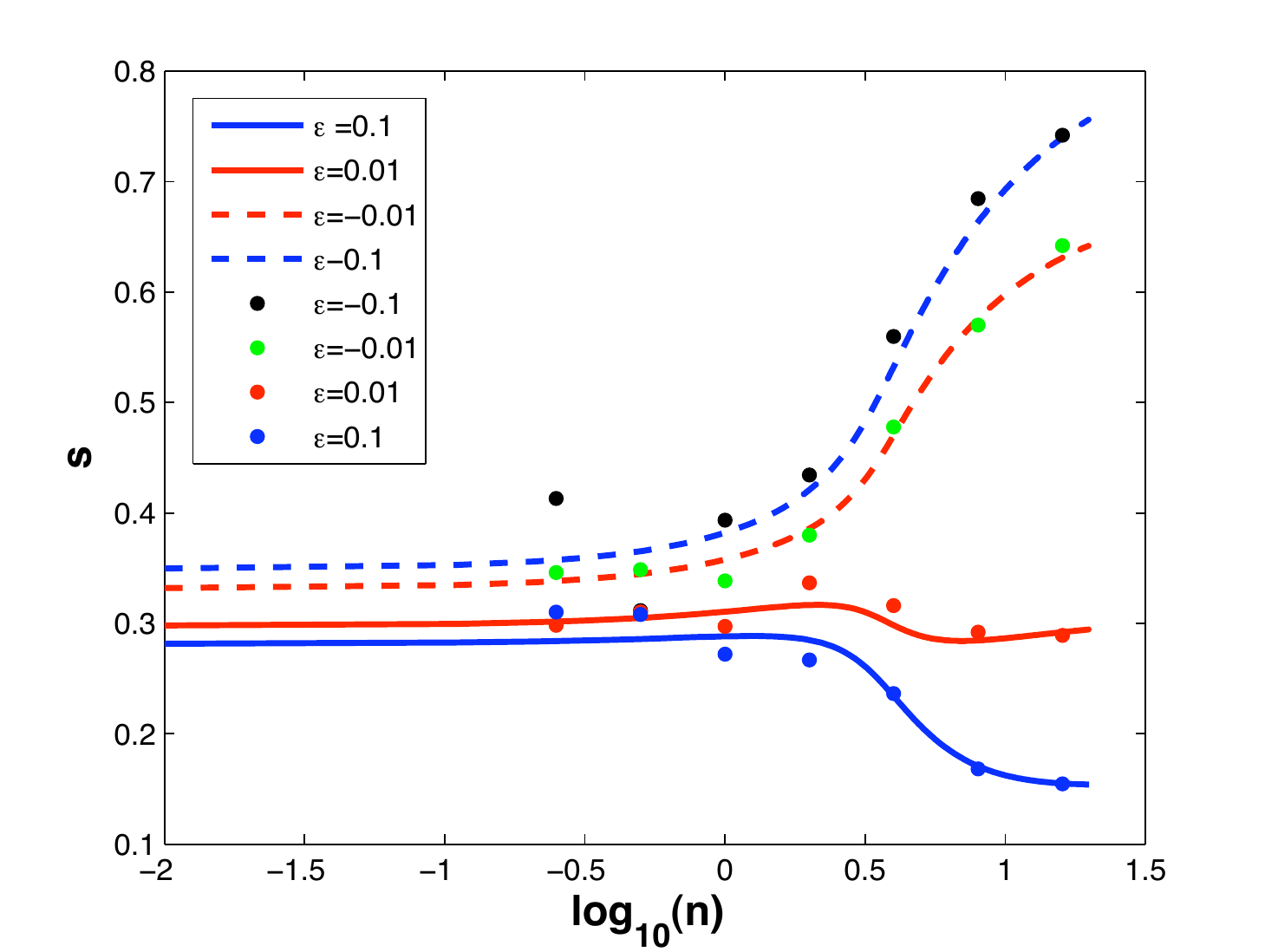}
\caption{\footnotesize {\textit{Supply as a function of $n$ for different values of $\overline{\epsilon}$, for $\sigma_\epsilon^2=0.01$ and $s_0=1$.} }}
\label{fig:eps1}
\end{center}
\end{figure}

The main features of this generalized problem are as follows:
\begin{enumerate}
\item  As for the homogeneous case, the fluctuations of returns eventually become very small as the market
complexity $n$ increases (see figure (\ref{fig:vol}) where we plot the volatility of returns
$\Sigma=\E_{\pi}[r-\E_{\pi}[r]^2]$). However, we can observe that the value $n^*$ beyond which the volatility approaches zero depends on the width of the risk premium distribution, and it increases with $\sigma_\epsilon^2$.

\item The sharp transition previously observed in the behavior of the average supply for large $n$ becomes
smooth as soon as the risk premium distribution has a finite width (see Figure (\ref{fig:eps1})).
\end{enumerate}

The sharpness of the crossover in the behavior of $s$ as a function of $\bar\epsilon$ is enhanced as the demand $s_0$ for derivatives increases. This signals a transition from a supply limited equilibrium, where the main determinant of the supply $s_i$ of derivatives is banks' profits, to a demand limited equilibrium, where $s_i$ is mostly limited by the finiteness of the investors' demand. In order to make this point explicit, it is instructive to investigate the case of unbounded supply ($s_0\to\infty$), because the transition becomes sharp in this limit.
The representative derivative, in this case, is described by
\begin{equation}
s_z=\max\left\{0,\frac{\sqrt{\frac{g+\Delta}{(1+\chi)^2}+\sigma_{\epsilon}}z-\overline{\epsilon}}{\nu}\right\},
\end{equation}
and, as before,
\begin{equation}
g=n\rm{E}_z[s_z^2],
\end{equation}
\begin{equation}
\chi=\frac{n\rm{E}_z[s_zz](1+\chi)}{\sqrt{g+\Delta+\sigma_{\epsilon}^2 (1+\chi)^2}}.
\end{equation}

\begin{figure}
\begin{center}
\includegraphics[width=10cm]{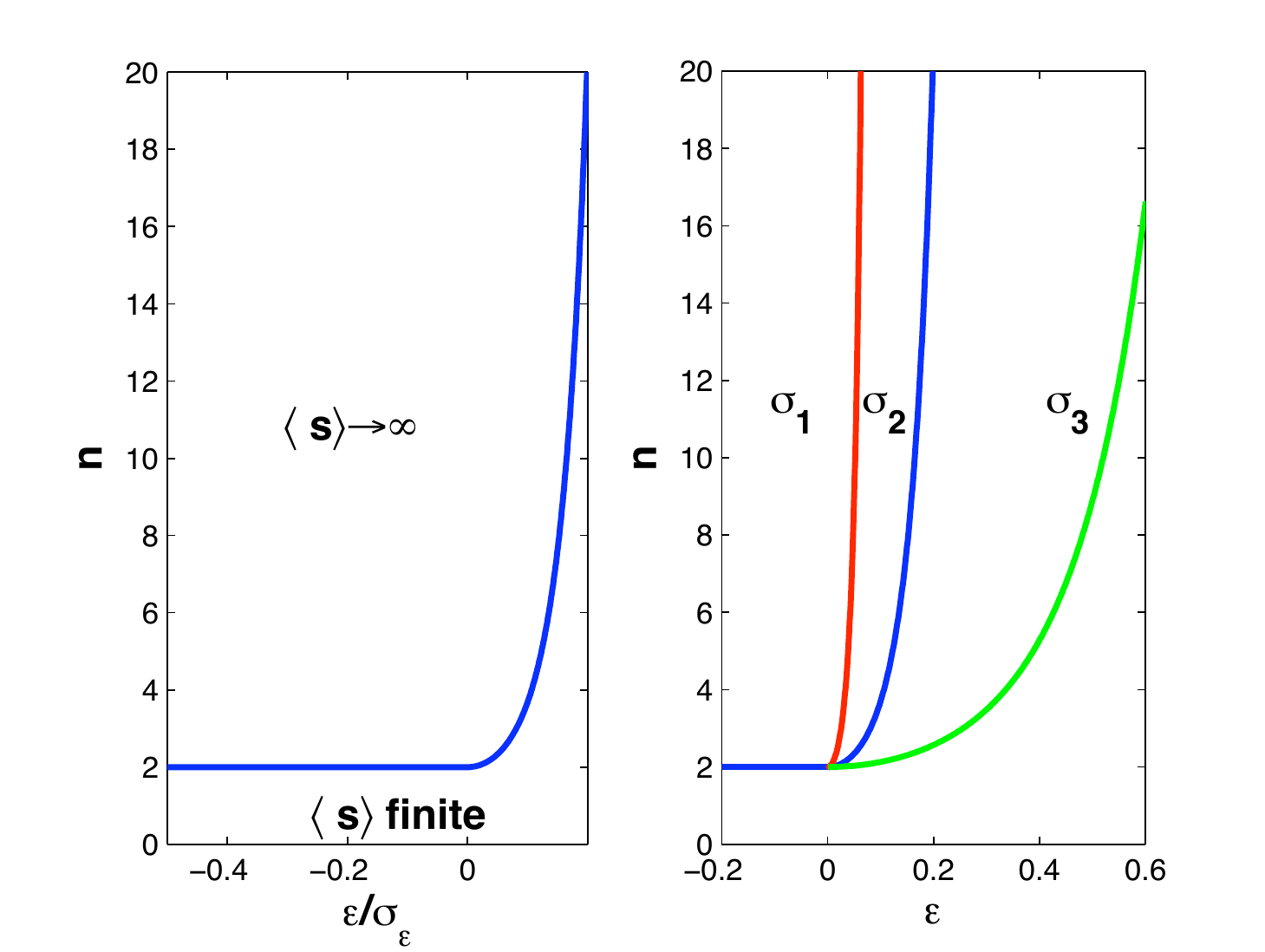}
\caption{\footnotesize {\textit{Phase diagram for the case of unbounded supply. Left panel: the plane is divided into two region. Above the blue line the supply diverges, while bleow it remains finite. Right panel: critical line for different values of $\sigma_{\epsilon}$ ($\sigma_3>\sigma_2>\sigma_1$) .} }}
\label{fig:phase_space}
\end{center}
\end{figure}
In contrast with the bounded case and similarly to the case $\sigma_{\epsilon}=0$, the system displays a phase separation in the $(\overline{\epsilon},n)$ plane between a region in which the average supply remains finite and a region in which the volume of the traded assets diverges (see left panel of Figure \ref{fig:phase_space}).

Introducing a distribution of risk premia with variance $\sigma_{\epsilon}$ has then very different effects depending on whether the supply is bounded or not. In the first case  $\sigma_{\epsilon}$ acts as a regularizer preventing the occurrence of a sharp phase transition. In the second case, namely $s_0\to\infty$, $\sigma_{\epsilon}\neq 0$ entails a deformation of the critical line that tends to flatten along the $n=2$ line as $\sigma_{\epsilon}$ grows (see right panel of Figure \ref{fig:phase_space} ).
This characterization allows to get some insights  on the case in which prices are dynamically generated.\\
As we showed before, in such a situation the system is characterized by a distribution of effective risk premia which becomes broader as market increases in complexity. Hence, upon increasing $n$, one expects a transition from a situation where volumes are limited by the profitability of derivative trading, to one where the demand is saturated.

\section{Conclusions}

We presented here a very stylized description of an arbitrage free market,  showing that  also whithin the ideal context of APT  the uncontrolled proliferation of financial intruments can lead to large fluctuations and instability in the market. Interestingly, within our simplified model, the proliferation of financial instruments drives the market through a transition to a state where volumes of trading rapidly expand and saturate investors' demand.

In order for these kind of models to be more realistic, some improvements certainly are needed.
 For instance, the demand for derivatives can be derived from a state dependent utility function for consumption at $t=1$.  This could also be a way to obtain an endogenous risk neutral measure, one of the main problems in the picture of dynamical prices considered above being that the risk neutral measure was fixed  exogenously.
We believe that, while accounting for these effects can make the model more appealing, the collective behavior of the model will not significantly change. Indeed, the qualitative behavior discussed above is typical of a broad class of models \cite{JPAreview} and mainly depends on the proliferation of degeneracies in the equilibria of the model.

The focus of the present paper is on theoretical concepts. Their relevance for real markets requires quantitative estimates of the parameters. Given the abstract nature of the model, this appears to be a non-trivial task which is an interesting avenue of future research.

It has been recently suggested that market stability appears to have the properties of a public good \cite{ShinMorris}. A public good is a good {\em i)} whose consumption by one individual does not reduce its availability for consumption by others (non-rivalry) and {\em ii)} such that no one can be effectively excluded from using the good (non-excludability). At the level of the present stylized description, the expansion in the repertoire of traded assets introduces an externality which drives the market to unstable states. This suggests that systemic instability may be prevented by the introduction of a tax on derivative markets, such as that advocated long ago  for foreign exchange markets by Tobin \cite{TobinTax}, or by the introduction of  "trading permits", similar to those adopted to limit Carbon emissions \cite{carbon}.
The stabilizing effect of a Tobin tax has already been shown within a model of a dynamic market which is mathematically equivalent to the one presented here \cite{Tobin}.

In summary, this paper suggests that  the ideal view of the markets on which financial engineering is based is not compatible with market stability. The proliferation of financial instruments makes the market look more and more similar to an ideal arbitrage-free, efficient and complete market. But this occurs at the expense of market stability. This is reminiscent of the instability
discussed long ago by Sir Robert May \cite{may} which develops in ecosystems upon increasing bio-diversity\footnote{This similarity can be made even more precise, within the model ecosystems discussed in \cite{JPAreview}.}. For ecologies this result is only apparently paradoxical. Indeed the species which populate an ecosystem can hardly be thought of as being drawn at random, but are rather subject to natural selection. Indeed, on evolutionary time scales stability can be reconciled with bio-diversity, as shown e.g. in Ref. \cite{Jain}. The diversity in the ecosystem of financial instruments has, by contrast, been increasing at a rate much faster than that at which selective forces likely operate.

In contrast with the axiomatic equilibrium picture of APT, on which financial engineering is based, the model discussed here provides a coherent, though stylized, picture of a financial market  as a system with interacting units. In this picture, concepts such as no-arbitrage, perfect competition, market efficiency or completeness arise as {emergent} properties of the aggregate behavior, rather than being postulated from the outset. We believe that such an approach can potentially shed light on the causes of and conditions under which liquidity crises, arbitrages and market crashes occur.

\section{Acknowledgements}
MM gratefully acknowledge discussions with J.-P. Bouchaud, D. Challet, A. Consiglio, M. Dacorogna, I. Kondor and M. Virasoro. Interaction with C. Hommes, specially in relation to  Ref. \cite{Hommes}, has been very inspiring. This work was supported by the ComplexMarkets E.U. STREP project 516446 under FP6-2003-NEST-PATH-1.

\appendix
\section{Calculation}
The problem is then the one of finding the ground state of the Hamiltonian
\begin{equation}
H=\frac{1}{2}\sum_{\omega}\pi^{\omega}(r^{\omega})^2+\sum_i\frac{\epsilon_i}{\Omega} s_i=\frac{1}{2}\sum_{\omega=1}^\omega {r^\omega}^2+\sum_{i=1}^Ng(s_i),
\end{equation}
is carried out in the usual manner. Let's write down the partition
function
\begin{eqnarray}
Z_{q,a} & = & {\rm Tr}_{r,s} e^{-\beta H}\delta\left(\sum_\omega q^
\omega r^\omega\right)\prod_{\omega =1}^\Omega
\delta\left(r^\omega-d^\omega-\sum_is_i  a_i^\omega\right)\\
& = & \int \!\frac{du}{2\pi}
{\rm Tr}_{r,s} e^{-\beta\sum_ig(s_i)}\prod_{\omega =1}^\Omega
e^{iur^\omega q^\omega-\frac{\beta}{2}{r^\omega}^2}\delta\left(r^
\omega-d^\omega-\sum_is_i  a_i^\omega\right)\\
  & =&  \int \!\frac{du}{2\pi}
{\rm Tr}_{r,s,\xi} e^{-\beta\sum_ig(s_i)}\prod_{\omega =1}^\Omega e^{iur^\omega
q^\omega-\frac{\beta}{2}{r^\omega}^2 +ir^\omega \xi^\omega-id^\omega \xi^\omega-i \xi^\omega\sum_i s_i
 a_i^\omega }
\end{eqnarray}
where we have used the shorthand ${\rm Tr}$ to indicate integrals on the variables in the index. Here it is
understood that all variables $s_i$ are integrated from $0$ to $1$ and all variables $r^\omega$ and $\xi^\omega$
are integrated over all the real axis, with a factor $1/ (2\pi)$ for each $\omega$. The next step is to
replicate this, i.e. to write

\begin{eqnarray*}
Z^m_{q,a} & = &  \int \!\frac{d\vec{u}}{2\pi} {\rm Tr}_{\vec{r},\vec{s},\vec{\xi}}
e^{-\beta\sum_{i,a}g(s_{i,a})} \prod_{\omega =1}^\Omega e^{i\sum_a [u_a r^\omega_a
q^\omega-\frac{\beta}{2}{r^\omega}^2_a +ir^\omega_a \xi^\omega_a-id^\omega \xi^\omega_a]-i \sum_i  a_i^\omega
\sum_a  s_{i,a}\xi^\omega_a
} \\
  & = &  \int \!\frac{d\vec{u}}{2\pi}
{\rm Tr}_{\vec{s},\vec{\xi}} e^{-\beta\sum_{i,a}g(s_{i,a})}\prod_{\omega =1}^\Omega
e^{-\frac{1}{2\beta}\sum_a\left(u_aq^\omega+ \xi_a^\omega\right)^2-id^\omega \sum_a \xi^\omega_a -i\sum_i
 a_i^\omega \sum_a  s_{i,a}\xi^\omega_a }
\end{eqnarray*}
where the sums on $a$ runs over the $m$ replicas.
In the second equation above, we have performed the
integrals over $r_a^\omega$.
We can now perform the average over the random variables $a_i^\omega$
which will be assumed to be normal with zero average and variance
$1/\Omega$. This yields

\begin{eqnarray}
\label{Zqm}   \langle Z^m_{q}\rangle  &= &
\int d \mathbf{G}\int \!\frac{d\vec{u}}{2\pi}
{\rm Tr}_{\vec{\xi}} \prod_{\omega =1}^\Omega
e^{-\frac{1}{2\beta}\sum_a\left(u_aq^\omega+\xi_a^\omega
\right)^2-id^\omega \sum_a \xi^\omega_a
-\frac{1}{2}\sum_{a,b}\xi_a^\omega \xi_b^\omega G_{a,b}}\langle I_s(\mathbf{G})\rangle_{\epsilon}\\
\langle I_s(\mathbf{G})\rangle_{\epsilon} & = & \Big\langle{\rm Tr}_{\vec{s}} e^{-\beta\sum_{i,a}g(s_{i,a})}
\prod_{a\le b}\delta\left(\Omega G_{a,b}-\sum_is_{i,a}s_{i,b}\right)\Big\rangle_{\epsilon},
\end{eqnarray}
where the average $\langle\ldots\rangle_{\epsilon}$ is taken over gaussian variables with mean $\overline{\epsilon}$ and variance $\sigma_{\epsilon}^2$ and the symbol $d\mathbf{G}$ stands for integration over all the independent entries of the matrix $\mathbf{G}$.
In order to evaluate the latter we use a standard delta function identity
bringing into play the matrix $\mathbf{R}=R_{a,b}$ conjugated to the overlap matrix $\mathbf{G}=G_{a,b}$.
The $\epsilon$-average is evaluated as follows
\begin{eqnarray*}
\langle I_s(\mathbf{G})\rangle_{\epsilon} & = &  \int \!d \mathbf{R}
{\rm Tr}_{\vec{s}} e^{\sum_{a\le b} R_{ab}
[\Omega G_{ab}-\sum_is_{i,a}s_{i,b}]}\int_{-\infty}^{+\infty}\prod_i \left[d\epsilon_i e^{- \frac{(\epsilon_i-\overline{\epsilon})^2}{2\sigma_{\epsilon}^2}-\beta\sum_{a}g(s_{i,a})} \right]\\
& =&  \int \!d \mathbf{R}
{\rm Tr}_{\vec{s}} e^{\sum_{a\le b} R_{ab}
[\Omega G_{ab}-\sum_is_{i,a}s_{i,b}] -\beta\sum_{i,a}\overline{\epsilon} s_i^a+\frac{\Omega \beta^2\sigma_{\epsilon}^2}{2}\sum_{a,b} G_{a,b}},
\end{eqnarray*}
where the quadratic term $(\sum_a s_i^a)^2$ arising from the gaussian integration has been replaced by the overlap matrix.
Taking the replica symmetric (RS) ansatz for $\mathbf{G}$ and $\mathbf{R}$
\begin{equation}
\label{RS}
G_{ab}=g+\frac{\chi}{\beta}\delta_{a,b},~~~~~~R_{ab}=-\beta^2 r^2+
\frac{\beta^2 r^2+\beta\nu}{2}\delta_{a,b}
\end{equation}
the $\vec{s}$-independent part of the exponent in the integrand takes the form
\begin{eqnarray*}
& \Omega\sum_{a\leq b}  R_{a,b} G_{a,b}+\frac{\Omega \beta^2 \sigma_{\epsilon}^2}{2}\sum_{a,b} G_{a,b}\\
& =  \Omega m\left(g+\frac{\chi}{\beta}\right)\left(\frac{\beta\nu}{2}-\frac{\beta^2r^2}{2}\right)-\frac{\Omega m(m-1)}{2}g\beta^2 r^2+\frac{\Omega\sigma_{\epsilon}^2}{2}(m^2\beta^2 g+m\beta\chi)\\
& \simeq \frac{\Omega m\beta}{2}\left(\nu g-r^2 \chi+\sigma_{\epsilon}^2\chi +\nu\chi/\beta\right)
\end{eqnarray*}
where we have neglected terms of order $m^2$ in view of the $m\to 0$ limit.
This yields
\begin{equation} \label{IofG}
 I_s(\mathbf{G})=\int \!d\mathbf{R}
e^{\frac{\Omega m\beta}{2}[\nu g-r^2\chi+\sigma_{\epsilon}^2\chi+\nu\chi/\beta]} W_{\overline{\epsilon}}[\nu,r]^N,
\end{equation}
where we have defined
\begin{equation}
W_{\overline{\epsilon}}[\nu,r]=\left[{\rm
Tr}_{\vec{s}} e^{-\beta\sum_{a}\left[\overline{\epsilon} s_a+\frac{\nu}{2}s^2_a\right]
+\frac{1}{2}\left(\beta r\sum_a s_a\right)^2}\right]^N.
\end{equation}
In the limit $m\to 0$ this quantity can be evaluated as follows
\begin{eqnarray*}
W_{\overline{\epsilon}}[\nu,r] & = e^{N \log \left[ {\rm
Tr}_{\vec{s}} e^{-\beta\sum_{a}\left[\overline{\epsilon} s_a+\frac{\nu}{2}s^2_a\right]
+\frac{1}{2}\left(\beta r\sum_a s_a\right)^2}\right]}\\
&=  \exp\left\{N \log \left[ {\rm
Tr}_{\vec{s}} e^{-\beta\sum_{a}\left[\overline{\epsilon} s_a+\frac{\nu}{2}s^2_a\right]}
\Big\langle e^{z\beta r \sum_a s_a}\Big\rangle_z \right] \right\}
\end{eqnarray*}
where we have performed a Hubbard-Stratonovich transformation in order to decouple the $\{s^a\}$ variables introducing the average $\langle\ldots\rangle_z$ over the gaussian variable $z$.
Clearly:
\begin{equation}
W_{\overline{\epsilon}}[\nu,r] =\exp\left\{N\log\Big\langle\left(\int_0^1 ds e^{-\beta [s^2+s(\overline{\epsilon}-zr)]}\right)^m\Big\rangle_z\right\}.
\end{equation}
Exploiting the usual identity
\begin{equation}
\log\langle X^m \rangle\simeq m\langle\log X\rangle,
\end{equation}
valid for $m\to 0$,
we finally obtain
\begin{equation}\label{W}
W_{\overline{\epsilon}}[\nu,r] =\exp\left\{N m\Big\langle\log\int_0^1 ds e^{-\beta [s^2+s(\overline{\epsilon}-zr)]}\Big\rangle_z\right\}.
\end{equation}
Inserting (\ref{W}) into (\ref{IofG}) we eventually get
\begin{equation} \label{IofGfinal}
 I_s(\mathbf{G})=\int \!d\mathbf{R}
e^{\Omega m\beta\left\{\frac{1}{2}[\nu g-r^2\chi+\sigma {\epsilon}^2\chi+\nu\chi/\beta]
+\frac{n}{\beta}\left\langle\log\int_0^\infty\!ds e^{-\beta[\overline{\epsilon} s+\nu
s^2/2-rsz]}
\right\rangle_z\right\}}
\end{equation}
After inserting (\ref{IofGfinal}) into (\ref{Zqm}) we have to perform the $(m\cdot \Omega)$ integrals in $\xi_a^ \omega$.
For each $\omega$
\begin{eqnarray}\label{J}
J[\chi,g,u]&=\int\!d\vec{\xi} e^{-\frac{1}{2\beta}\sum_a\left(u_aq^\omega+\xi_a\right)^2-id^\omega \sum_a \xi_a
-\frac{1}{2}\sum_{a,b}\xi_a \xi_b G_{a,b}}
\end{eqnarray}
The matrix $\mathbf{G}$, in the RS ansatz has two distinct eigenvalues:
\begin{eqnarray}
 a_{\parallel}&= mg+\frac{\chi}{\beta}&\qquad\mbox{multiplicity } 1\\
 a_{\perp}&=\frac{\chi}{\beta}&\qquad\mbox{multiplicity } m-1.
\end{eqnarray}
Therefore the determinant of $\left(\mathbf{G}+\frac{\mathbf{I}}{\beta}\right)$ is clearly
\begin{equation}\label{det}
\det\left(\mathbf{G}+\frac{\mathbf{I}}{\beta}\right)=e^{-\frac{1}{2}\left[m\log(\chi+1)+\log\left(1+\frac{\beta m g}{1+\chi}\right)\right]}.
\end{equation}
The expression (\ref{det}) assists performing the gaussian integral in (\ref{J}) as
\begin{equation}
J[\chi,g,u]=e^{-\frac{1}{2}\log\left(1+\beta m\frac{g}{1+\chi}\right)-\frac{m}
{2}\log(1+\chi)
+\frac{\beta m}{2(1+\chi+m\beta g)}( uq^\omega
+id^\omega)^2-\frac{\beta m}{2}( uq^\omega)^2}
\end{equation}
where we have rescaled $u$ as $u\to\beta u$ and considered for $u^a$ the form $u^a=u$ $\forall a$.
Taking the average over $q^\omega$ in the limit $m\to 0$ we use the
fact that
\[
\langle e^{m f(q)}\rangle_q\simeq 1+m\langle f(q)\rangle_q
\simeq e^{m\langle f(q)\rangle_q}
\]
which in our case yields
\[
\langle f(q)\rangle_q=-\frac{\beta\chi}{1+\chi}u^2-\frac{1}{2}\frac{\beta \langle d^2\rangle}{1+\chi},
\]
having taken for $q$ an exponential distribution with average $1$ .
Hence
\begin{eqnarray}
\langle Z^m_{q}\rangle  & = &
\int d \mathbf{G}\int d \mathbf{R}\int \!\frac{d\vec{u}}{2\pi} e^{m\Omega\beta F}\\
\label{Fofbeta}F & = &
-\frac{u^2\chi}{1+\chi}-\frac{1}{2}\frac{g+\langle
d^2\rangle}{1+\chi}-\frac{1}{2\beta}\log(1+\chi)+\frac{1}{2}\left[\nu g-r^2\chi+\sigma_{\epsilon}^2\chi+\frac{\nu\chi}{\beta}\right]
\nonumber\\
& ~ & +\frac{n}{\beta}\left\langle\log\int_0^\infty\!ds e^{-\beta[\overline{\epsilon} s+\nu s^2/2-rsz]}
\right\rangle_z.
\end{eqnarray}
We first observe that  the saddle point equation for $u$
yields $u=0$. Then we take the limit $\beta\to\infty$ which gives

\begin{equation}
\label{Ffin}
F =
-\frac{1}{2}\frac{g+
\langle d^2\rangle}{1+\chi}+\frac{1}{2}\left[\nu g-r^2\chi+\sigma_{\epsilon}^2\chi \right] -n
\left\langle\min_{s\ge 0}
\left[\overline{\epsilon} s+\frac{\nu}{2} s^2-rsz\right]
\right\rangle_z
\end{equation}

The saddle points equations, obtained differentiating (\ref{Fofbeta}) with respect to the order parameters and  sending $\beta$ to $\infty$, read
\begin{eqnarray}
\label{r} r^2 & = &\frac{g+\langle d^2\rangle}{(1+\chi)^2}+\sigma_{\epsilon}^2\\
\label{nu} \nu & = &\frac{1}{1+\chi}\\
\label{g} g &=& n\langle s_z ^2\rangle_z\\
\label{chi} r\chi & =& n\langle s_z z\rangle_z\\
\end{eqnarray}
where $s_z=\min\{s_0,\max\{0,(rz-\overline{\epsilon})/\nu\}\}$, since,  in our case,  the supply is limited to
$0\le s\le s_0$. The above set of equations can then be recasted in the form
\begin{eqnarray}
s_z &=\min\left\{1,\max\left\{0,\left(z\sqrt{\frac{g+\langle d^2\rangle}{(1+\chi)^2}+\sigma_{\epsilon}^2}-\overline{\epsilon}\right)(1+\chi)\right\}\right\}\\
g &=n\E_z[s_z^2]\\
\chi &=\frac{n\E_z[s_zz](1+\chi)}{\sqrt{g+\Delta+\sigma_{\epsilon}^2(1+\chi)^2}}.
\end{eqnarray}
The above calculation can also be performed, in order to probe the solution, introducing an auxiliary field $h^{\omega}$ coupled to the returns $r^{\omega}$. This allows also to easily compute the average and the fluctuations of returns by deriving with respect to $h$ the logarithm of the free energy and then setting $h=0$:
 \begin{eqnarray}
\bar r & = & \sum_\omega \pi^\omega r^\omega = \frac{\bar d}{1+\chi}\\
\delta r^2 & = & \sum_\omega \pi^\omega (r^\omega-\bar r)^2 =\frac{g+\langle d^2\rangle -\langle d\rangle^2}{(1+\chi)^2}
\end{eqnarray}

\section{Critical line}
We show here how it is possible to find the critical line. Let us consider the case $s\in[o, \infty)$, so that $s_z={\rm max}\left\{0,\frac{r}{\nu}(z-z_0)\right\}$, with $z_0=\frac{\overline{\epsilon}}{r}$.
From the saddle pont equations we get
\begin{equation}
g=\frac{n[\langle d^2\rangle+\sigma_{\epsilon}^2(1+\chi)^2]I_2(z_0)}{1-nI_2(z_0)}
\end{equation}
where we have defined $I_2(z_0)=\int_0^\infty dz \frac{e^{-z^2/2}}{\sqrt{2\pi}}(z-z_0)^2$.
Inserting now this expression for $g$ into equation (\ref{r}) and explointing $r=\overline{\epsilon}/z_0$ we obtain
\begin{equation}
\frac{\overline{\epsilon}^2}{z_0^2}(1-nI_2(z_0))=\sigma_{\epsilon}^2+\frac{\langle d^2\rangle}{(1+\chi)^2}
\end{equation}
that using (\ref{chi}) can be writen as
\begin{equation}
\frac{\overline{\epsilon}^2}{z_0^2}(1-nI_2(z_0))=\sigma_{\epsilon}^2+\langle d^2\rangle (1-nI_1(z_0))^2,
\end{equation}
where $I_1=\int_0^\infty dz \frac{e^{-z^2/2}}{\sqrt{2\pi}}(z-z_0) z$.
If we now look for a solution in the case $\chi\to\infty$ the above equation reduces to
\begin{equation}
\frac{\overline{\epsilon}^2}{\sigma_{\epsilon}^2}\left(1-\frac{I_2(z_0)}{I_1(z_0)}\right)=z_0^2
\end{equation}
and we also have that $n=1/I_1(z_0)$.
These equations define the critical line. In particular it is clear at this level that the dependence on the parameters of the risk premia distribution enters trough the ratio $\overline{\epsilon}/\sigma_{\epsilon}$

\end{document}